\documentclass[fleqn,twoside]{article}
\usepackage{espcrc2}

\newcommand{\beq}{\begin{equation}}
\newcommand{\eeq}{\end{equation}}

\def\dg{\Delta \Gamma_d}
\def\bea{\begin{eqnarray}}
\def\eea{\end{eqnarray}}
\def\nn{\nonumber}
\def\barr{\begin{eqnarray}}
\def\earr{\end{eqnarray}}

\def\lsim{\raise0.3ex\hbox{$\;<$\kern-0.75em\raise-1.1ex\hbox{$\sim\;$}}}
\def\gsim{\raise0.3ex\hbox{$\;>$\kern-0.75em\raise-1.1ex\hbox{$\sim\;$}}}

\def\gh{\Gamma_H}
\def\gl{\Gamma_L}
\def\dg{\Delta \Gamma_d}

\title{Measurement of the Width Difference of $B_d$ Mesons
\thanks{Presented by T.Yoshikawa at KEKTC5.}}

\author{A.S. Dighe\address{
   Max-Planck-Institute for Physics, F\"ohringer Ring 6,
        D-80805 Munich, Germany},               
        T. Hurth\address{
        CERN, Theory Division, CH--1211 Geneva 23, Switzerland}, 
        C.S. Kim\address{
        Department of Physics and IPAP, Yonsei University, 
        Seoul 120-749, Korea     } and 
  T. Yoshikawa\address{
         Department of Physics and Astronomy, 
          University of North Carolina, \\
        Chapel Hill, NC 27599-3255, USA}}

\begin{document}

\begin{abstract}
We estimate $\dg/\Gamma_d$, including $1/m_b$ contributions
and part of the next-to-leading order QCD corrections, and find it to 
be around $0.3\%$.
We show the methods to measure $\dg/\Gamma_d$ by using at least 
two different final states on the untagged $B_d$ decay.  
The nonzero width difference can also be used to identify new physics 
effects and to resolve a twofold discrete ambiguity in the $B_d-\bar{B}_d$ 
mixing phase. 
With the high statistics and accurate time resolution of the upcoming
LHC experiment, the measurement of $\dg$ seems to be possible.
This measurement would be important for an accurate 
measurement of $\sin2\phi_1$ at the LHC. 
We also derive an upper bound on the value of $\dg/\Gamma_d$
in the presence of new physics.  
\end{abstract}

\maketitle
The neutral $B_d$ meson system has two mass eigenstates 
which have the mass difference and lifetime difference. 
The mass difference has been measured well but the lifetime difference 
has not been done because it is very tiny. 
Within the standard model (SM), the difference in the decay 
widths is CKM-suppressed with respect to 
that in the $B_s$ system. A rough estimate leads to
$\frac{\Delta \Gamma_d}{\Gamma_d} \, \sim \, 
\frac{\Delta \Gamma_s}{\Gamma_s} \cdot \lambda^2
\approx  0.5 \% ~~,$\, 
where $\lambda = 0.225$ is the sine of the Cabibbo angle, and we have taken 
$\Delta \Gamma_s/\Gamma_s \approx 15\%$ \cite{BBGLN} (see also \cite{Ben,Bec}). 
Here $\Gamma_{d(s)} = (\Gamma_L + \Gamma_H)/2$ is the average decay width 
of the  light and heavy $B_{d(s)}$ mesons ($B_L$ and $B_H$ respectively).
We denote these decay widths by $\Gamma_L,
\Gamma_H$ respectively,
and define $\Delta \Gamma_{d(s)} 
\equiv \gl - \gh$. 

At the present accuracy of measurements,
this lifetime difference $\dg$ can well be ignored. As a result,
the measurement and the phenomenology of $\dg$ have been neglected
so far, as compared with the lifetime difference in the $B_s$ system
for example. However,
with the possibility of experiments with high time resolution and 
high statistics, such as at the  LHC, 
this quantity is becoming more and more relevant.



With the possibility of experiments with high time
resolution and high statistics, it is worthwhile to have a look
at this quantity and make a realistic estimate of the 
possibility of its measurement (see also \cite{Paper,Workshop}).


In \cite{Paper} we estimated $\dg/\Gamma_d$ including $1/m_b$ contributions
and part of the next-to-leading order QCD corrections.
We find that adding the latter corrections decreases 
the value of $\dg/\Gamma_d$ computed at the leading order by 
a factor of almost $2$. The final result is
\begin{equation}
\Delta \Gamma_d / \Gamma_d = (2.6 ^{+1.2}_{-1.6}) \times 10^{-3} \, .
\end{equation}
The two mass eigenstates of the neutral $B_d$ system 
have slightly different lifetimes.
Using another expansion of the partial NLO QCD corrections proposed
in \cite{breport}, we get
\begin{equation}
\Delta \Gamma_d / \Gamma_d = (3.0 ^{+0.9}_{-1.4}) \times 10^{-3}\, ,
\end{equation}
where we have used the preliminary values 
for the bag factors from the JLQCD collaboration~\cite{jlqcd}. 
In the error estimation, the errors are
the uncertainties on the values of
the CKM parameters, of the bag parameters, of the mass of the $b$ quark,
and of the measured value of $x_d$.
Further sources of error are the assumption of naive factorization 
made for  the $1/m_b$ matrix elements, 
the scale dependence  and the missing  terms in the NLO contribution. 
Although the latter error is decreased 
in the second estimate by smallness of  CKM factors, 
a complete NLO calculation is definitely desirable for the 
result to be more reliable. 


The most obvious way of trying to measure the width difference is 
through the semileptonic decays, however we can not extract the quantity 
which is linear in  $\Delta \Gamma_d / \Gamma_d$. The time measurements
of an untagged $B_d$ decay to single final state is sensitive only to 
quadratic terms in $\Delta \Gamma_d / \Gamma_d$. So this method would 
involve measuring a quantity as small 
as $(\Delta \Gamma_d / \Gamma_d)^2 \sim 10^{-5}$, which is too small 
to measure. 

However, combining time measurements from two different final states
can enable us to measure quantities linear in $\Delta \Gamma_d / \Gamma_d$. 
Indeed, we can measure the ratio of two untagged 
lifetimes for two final states:
\bea
\frac{\tau_{b_1}}{\tau_{b_2}}=1+\frac{b_2-b_1}{2}
       \frac{\Delta\Gamma_d }{\Gamma_d} 
     + {\cal O} \left[ (\dg/\Gamma_d)^2 \right],
\eea 
where the $b$ is the quantity depend on the final state. 
This indicates the necessity of at least two different final states 
to extract $\Delta \Gamma_d / \Gamma_d$. 
 
A viable option, perhaps the most efficient among the ones 
considered in \cite{Paper}, is to compare the measurements of the
untagged lifetimes  of the semileptonic decay mode $\tau_{SL}$ and 
of the CP-specific decay modes $\tau_{CP_{\pm}}$. For each final states, 
$b_{SL}=0$ and $b_{CP\pm}=\pm\cos(2\phi_1)$. 
The ratio between the two lifetimes 
$\tau_{CP\pm}$ and $\tau_{SL}$ is 
\begin{equation}
\frac{\tau_{SL}}{\tau_{CP\pm}} = 1 \pm \frac{\cos(2\phi_1)}{2} 
\frac{\dg}{\Gamma_d}
+ {\cal O} \left[ (\dg/\Gamma_d)^2 \right]
~~.
\label{cp-sl}
\end{equation}
The measurement of these two lifetimes 
should be able to give us a value of $|\dg|$,
since $|\cos(2\phi_1)|$ will already be known to a good accuracy
by that time.   

Since the CP-specific decay modes of $B_d$ 
(e.g. $J/\psi K_{S(L)}, D^+ D^-$) have smaller
branching ratios than the semileptonic modes, and 
the semileptonic data sample may be enhanced 
by including the self-tagging decay modes (e.g. $D_s^{(*)+}D^{(*)-}$)
 which also have large branching ratios,
we expect that the most useful combination will be the
measurement of $\tau_{SL}$ through all self-tagging decays and
that of $\tau_{CP_+}$ through the decay $B_d \to J/\psi K_S$.
After 5 years of LHC running, 
we should have about $5 \times 10^5$ events of 
$J/\psi K_S$ (\cite{lhc} table 3), 
whereas the number of semileptonic decays,   
at LHCb alone, that will be directly useful in the lifetime 
measurements is expected to be more than $10^6$ per year, 
even with conservative estimates of efficiencies.

At LHCb, the proper time resolution is expected to be
as good as $\Delta \tau \approx 0.03$ ps.
This indeed is a very small fraction of the $B_d$ lifetime
($\tau_{B_d} \approx 1.5$ ps \cite{pdg}), 
so the time resolution
is not a limiting factor in the accuracy of the measurement,
 and  the statistical error plays the dominant role.
Taking into account the estimated
number of $B_d$ produced 
the measurement of the
lifetime difference does not look too hard at first glance.
One may infer that if the number of relevant events
with the proper time of decay measured with the
precision $\Delta \tau$ is $N$, then the value
of $\Delta \Gamma_d / \Gamma_d$ is measured with an
accuracy of $1 / \sqrt{N}$. With a
sufficiently large number of events $N$, it should be
possible to reach the accuracy of 0.5\%
or better. 

We also point out the interlinked nature of the accurate measurements of 
$\phi_1$ and $\Delta \Gamma_d / \Gamma_d $ through the conventional 
gold-plated decay\cite{Paper,Proc3}. 
In the future experiments that aim to measure $\phi_1 $
to an accuracy of 0.005 or better, the corrections due to 
$\Delta \Gamma_d $ will form a major part of the systematic error, 
which can be taken care of by a simultaneous fit to 
$\sin(2\phi_1)$,$\Delta\Gamma_d$
and an effective parameter $\bar{\epsilon}$ that comes from the CP
violation in $K-\bar{K}$ and $B-\bar{B}$ systems, and also takes care of
small theoretical uncertainties.


The calculations of the width difference in $B_d$ 
and in the $B_s$ system (as in \cite{BBGLN}) run along
similar lines. However, there are some subtle differences involved,
due to the values of the different CKM elements involved, which
have significant consequences. 
In particular, whereas the upper
bound on the value of $\Delta\Gamma_s$ (including the effects of
new physics) is the value of $\Delta\Gamma_s({\rm SM})$ \cite{grossman},
the upper bound on $\Delta\Gamma_d$ involves a multiplicative
factor in addition to $\Delta\Gamma_d({\rm SM})$.
Using the definitions 
$\Theta_q \equiv {\rm Arg}(\Gamma_{21})_q , 
\Phi_q \equiv {\rm Arg}(M_{21})_q $,
where $q \in \{ d,s \}$, we can write 
\begin{equation}
\Delta\Gamma_q = - 2 |\Gamma_{21}|_q  \cos(\Theta_q - \Phi_q)~~.
\label{theta-phi}
\end{equation}
Since the contribution to $\Gamma_{21}$ comes only from tree
diagrams, we expect the effect of new physics on this quantity 
to be very small. We therefore take
$|\Gamma_{21}|_q$ and $\Theta_q$ to be unaffected by new physics. 
On the other hand, the mixing phase $\Phi_q$ appears from loop
diagrams and can therefore be very sensitive to new physics.
The effect of new physics on $\Delta\Gamma_s $ can be bounded 
by giving an upper bound on $\Delta\Gamma_s $:  
\begin{equation}
\Delta\Gamma_s \leq \frac{\Delta \Gamma_s({\rm SM})}{\cos(2\Delta\gamma)}
\approx \Delta \Gamma_s({\rm SM})~~,
\label{dgs-bd}
\end{equation}
with $ 2\Delta\gamma = - Arg[(V_{cb}^*V_{cs})^2/(V_{tb}^*V_{ts})^2]
\approx - 0.03 $.
Thus, the value of $\Delta \Gamma_s$ can only decrease in the
presence of new physics\cite{grossman}.

In the $B_d$ system, an upper bound for $\dg$,  
based on the additional assumption of three-generation
unitarity, can be derived: 
\begin{equation}
\Delta \Gamma_d \leq  \frac{\Delta \Gamma_d({\rm SM})}
{\cos[ {\rm Arg}(1 + \delta f) ]}~~.
\label{dgd-bd}
\end{equation}
We can calculate the
bound (\ref{dgd-bd}) in terms of the extent of the higher order NLO
corrections. In \cite{Paper}, we got $|{\rm Arg}(1 + \delta f)| < 0.6$,
so that we have the bound  
$\Delta \Gamma_d < 1.2 ~ \Delta \Gamma_d({\rm SM})$. 
A complete NLO calculation will be able to give a
stronger bound.

We have seen that the ratio of 
two effective lifetimes can enable us to measure the quantity
$\Delta\Gamma_{obs(d)} \equiv \cos(2\phi_1) \Delta\Gamma_d/\Gamma_d$.
In the presence of new physics, this quantity is in fact
(see eq.~(\ref{theta-phi}))
$\Delta\Gamma_{obs(d)} =  - 2 (|\Gamma_{21}|_d/\Gamma_d)  
\cos(\Phi_d) \cos(\Theta_d - \Phi_d)$. In SM, we get 
\bea
\Delta\Gamma_{obs(d)}({\rm SM}) &=& 2 (|\Gamma_{21}|_d/\Gamma_d) 
\cos(2 \phi_1) \nn \\  
& & ~~~~~~~ \times \cos[ {\rm Arg}(1 + \delta f) ]~~.
\eea
If $|\delta f| < 1.0$, we have $\cos[{\rm Arg}(1+\delta f)] > 0$ 
(in fact, from the fit in \cite{ckm-fit} and our error estimates, 
we have $\cos[{\rm Arg}(1+\delta f)] > 0.8$). 
Then $\Delta\Gamma_{obs(d)}({\rm SM})$ is predicted to be positive.
New physics is not expected to affect $\Theta_d$, but it may affect 
$\Phi_d$ in such a way as to make the combination 
$\cos(\Phi_d) \cos(\Theta_d - \Phi_d)$ change sign.
A negative sign of $\Delta\Gamma_{obs(d)}$ would  therefore 
be a clear signal of such new physics.


It is well known, that the 
$B_d$--$\bar{B}_d$ mixing phase $\Phi_d$ is efficiently
measured through the decay modes $J/\psi K_s$ and $J/\psi K_L$. 
If we take the new physics effects into account, the time-dependent
asymmetry is  
${\cal A}_{CP} = - \sin(\Delta M_d t) \sin(\Phi_d)$; 
in the SM, we have $\Phi_d = -2\phi_1$. 
The measurement of $\sin(\Phi_d)$ still allows for a discrete ambiguity
$\Phi_d \leftrightarrow \pi - \Phi_d$. 
It is clear that,  
if $\Theta_d$ can be determined  
independently of the mixing in the $B_d$ system,  
then measuring $\Delta\Gamma_{obs(d)}$,  which is proportional to 
$\cos(\Phi_d) \cos(\Theta_d - \Phi_d)$,  
resolves the discrete ambiguity in principle. 
We note that these features are unique to the $B_d$ system.

\section*{ACKNOWLEDGMENTS}
The work of C.S.K. was supported by Grant No. 2001-042-D00022 of the
KRF. The work of T.Y. was supported in part by the US Department of 
Energy under Grant No.DE-FG02-97ER-41036.

\end{document}